\documentclass[a4paper]{jpconf}
\usepackage{graphicx}
\usepackage{cite}
\usepackage{xcolor}
\usepackage{appendix}
\usepackage{amsmath,amssymb,amsfonts,latexsym,cancel}
\newcommand{\be}{\begin{equation}}
\newcommand{\ee}{\end{equation}}
\newcommand{\bea}{\begin{eqnarray}}
\newcommand{\eea}{\end{eqnarray}}

\begin{document}
\title{Instantaneous vacuum and States of Low Energy for a scalar field in cosmological backgrounds}

\author{Antonio Ferreiro$^{1,2}$ and Silvia Pla$^{3}$}
\address{$^{1}$Institute for Mathematics, Astrophysics and Particle Physics, Radboud University, 6525 AJ Nijmegen, The Netherlands
}
\address{$^{2}$Departamento de F\'isica Te\'orica, Universidad de Valencia and Instituto de F\'isica Corpuscular-CSIC/Universidad de Valencia, Burjassot–46100, Spain}

\address{$^3$Theoretical Particle Physics and Cosmology, King's College London, WC2R 2LS, UK}

\ead{ antonio.ferreiro@ru.nl}

\ead{silvia.pla\_garcia@kcl.ac.uk}

\begin{abstract}
We construct the instantaneous vacuum state for a quantum scalar field coupled to another classical scalar field as described in \cite{Ferreiro:2022hik}. We then compare it with the state of low energy constructed for a particular solution. We show that under physically motivated conditions they become very similar.  
\end{abstract}

\section{Introduction}

In our current cosmological model, quantum effects of matter play a key role in our understanding of the early Universe. In particular, during the inflationary period and the posterior reheating era generation of quantum fluctuations and particle production are ubiquitous. The best framework to study these effects is quantum field theory in curved spacetime, which in the last decades has been constructed in a mathematically consistent way. However, many of the relevant results have been established in a rather unpractical methodology for actual numerical calculations/simulations, which are fundamental in most cosmological scenarios.\\

One example of this is the consistent characterization of suitable vacuum states in terms of the so-called Hadamard condition \cite{Wald94,Fewster13}. This is a requirement on the singular behavior of the two-point correlation function at separate spacetime points. By satisfying this condition, the Wick polynomials of any degree can be guaranteed to exist, allowing the perturbative expansion of an interacting theory to be well-defined at all orders \cite{BF2000,hollands-wald-01, hollands-wald-02}. For the case of cosmological spacetimes, an equivalent notion has been proposed in terms of the behavior of the two-point function in the limit $k \to \infty$ \cite{parker-toms}. This condition, known as the adiabatic condition, has been shown to be equivalent to the Hadamard condition in the limit of infinite adiabatic order \cite{Fulling,Pirk93,Hollands01}. \\


 Obtaining a vacuum state that satisfies the Hadamard condition is, in general, a challenging task. However, in cosmological spacetimes one can successfully construct vacuum states that satisfy this condition by means of the low energy states, introduced for the first time in Ref. \cite{Olbermann:2007gn}. In this framework, the vacuum is defined as the state that minimizes the vacuum expectation value of the energy density averaged over a temporal window function supported around a well-motivated initial time $t=t_0$. This method has been successfully employed to obtain physically motivated vacua in bouncing inflationary scenarios in loop quantum cosmology \cite{Martin-Benito:2021szh}, in a CPT-symmetric universe \cite{Nadal23}, and for the Schwinger effect \cite{Alvarez-Dominguez:2023ten}. Furthermore, the prescription to construct those states was recently extended in Ref. \cite{Nadal23} for spin-$\frac{1}{2}$ fields. We note here that these low energy states are constructed starting from an analytic solution of the equation of motion of the scalar field modes. This is very restrictive for numerical calculations/simulations, where we wish to obtain numerical solutions given expanding parameter that may vary due to the backreaction of the produced quantum contributions of the scalar field. This issue justifies the search for different possibilities, that can be defined without the need of analytical solutions.\\

  In this context, we have recently proposed a possible vacuum state that could be constructed at any given time taking into account only the mode-by-mode subtraction terms of the regularized stress-energy tensor \cite{Ferreiro:2022hik}. For the subtraction terms, we used an extended version of the adiabatic regularization method, that includes an arbitrary scale parameter $\mu$. This was a generalization of previous work in which this vacuum state was computed for some relevant backgrounds but always using an on-shell type prescription for the adiabatic subtraction terms, which does not encapsulate the possible freedom of choice of the method \cite{Agullo:2014ica}. It turned out that it was not suitable for the case of an effective, time-dependent mass for the scalar field, which is the typical case when studying reheating models.  The state constructed with this method is known as the instantaneous vacuum and it is defined by requiring that the renormalized energy density and pressure vanish at a given initial time $t=t_0$ mode-by-mode. This vacuum state is, by definition, a state of fourth adiabatic order, which ensures a well-defined stress-energy tensor that can be then used to solve the semiclassical Einstein equations numerically.  \\

  This work aims to compare both methods for a quantized scalar field with a Yukawa interaction in a flat FLRW universe. We proceed as follows. In Section \ref{sec:Yukawa} we introduce the model and the extended adiabatic regularization method in this context. In Sections \ref{sec:IV} and \ref{sec:SLE} we briefly describe the instantaneous vacuum and the states of low energy respectively for the model under consideration. Finally, in Section \ref{sec:minkowski} we explicitly compare both methods. For simplicity, and since we need analytical solutions to build the states of low energy, we restrict ourselves to Minkowski spacetime. In section \ref{sec:conclusions} we summarise the main results of the proceeding.

\section{Quantum scalar field with a Yukawa interaction} \label{sec:Yukawa}
Consider an action for a massless, minimally coupled scalar field $X$, coupled to another classical massive and minimally coupled scalar field
\be S[X,\phi,g] = \frac12 \int \textrm{d}^4x \sqrt{-g} \left\{  g^{\mu \nu} \partial_{\mu} X \partial_{\nu} X -g\phi^2 X^2+  g^{\mu \nu} \partial_{\mu} \phi \partial_{\nu}\phi-M^2\phi^2 \right\} \ ,  \ee
where here we have both the interaction with the curvature and with a classical scalar field $\phi$. Here we will not take into account the quantization of the later, and we refer to \cite{Ferreiro:2022hik} for the discussion of the $\lambda \phi^4$ theory case. We also consider here a flat Friedman-Lemaitre-Robertson-Walker spacetime 
\be\textrm{d}s^2=a^2(\eta)(\textrm{d}\eta^2-\textrm{d}\vec x^2)\, ,
\ee
where $\eta$ is the conformal time coordinate. We also assume $\phi=\phi(\eta)$. This metric allow us to expand the field $X$ as
\bea
X(\vec{x}, \eta)=\frac{1}{(2 \pi)^3} \int \textrm{d}^3 k\left[A_{\vec{k}} h_{k}(\eta) e^{i \vec{k} \vec{x}}+A_{\vec{k}}^{\dagger} h_{k}^*(\eta) e^{-i \vec{k} \vec{x}}\right]\, .
\eea
The Klein-Gordon equation $\Box X +g \phi^2 X=0$ implies
\be \label{eq:mode-eq}
h_{k}^{\prime \prime}+2 \frac{a^{\prime}}{a} h_{k}^{\prime}+\left(k^2+g a^2 \phi^2 \right) h_{k}=0\, .
\ee
In an isotropic and homogeneous spacetime, the vacuum expectation value of the stress energy tensor takes the form 
\be
\langle T_{ab}\rangle=-g_{ab}\langle p\rangle+(\langle p\rangle+\langle \rho\rangle )u_a u_b\, , \label{Trhop}
\ee
where where $u^a$ is the unit vector normal to the homogeneous and isotropic hypersurfaces, and where the energy density and pressure are
\be
\langle \rho \rangle\equiv\frac{1}{(2\pi)^3}\int \text{d}^3k \langle \rho_k\rangle \,, \qquad  \langle p \rangle\equiv\frac{1}{(2\pi)^3}\int \text{d}^3k \langle p_k\rangle \,, 
\ee

\be
\begin{aligned}
 \label{rhop}
\langle\rho_k \rangle =\frac{1}{2a^2} \left(|h'_{k}|^2+\left(k^2+g \phi^2a^2\right)|h_{k}|^2\right)~~~
\langle p_k \rangle=\frac{1}{2a^2} &\left(|h'_{k}|^2-\left(\frac{k^2}{3}+g \phi^2a^2\right)|h_{k}|^2\right).
\end{aligned}
\ee
As it is well-known, these quantities are plagued with divergences and a regularization program is needed. We use here the adiabatic regularization method (see, for example, Ref. \cite{parker-toms}) in its extended version proposed in Ref. \cite{Ferreiro:2018oxx} and extended in Ref. \cite{Ferreiro:2022hik}.  This technique 
  eliminates the UV divergences by subtracting the adiabatic counterterms mode by mode under the $k$-integral, and includes an additional parameter $\mu$ to avoid the infrared divergences. After regularizing we obtain the finite quantities
\be
\langle\rho_k \rangle _{\textrm{ren}}=\langle\rho_k \rangle-C_\rho(\mu,k,\eta)\, , \qquad \langle p_k \rangle _{\textrm{ren}}=\langle p_k \rangle-C_p(\mu,k,\eta)\,.
\ee
where the coefficients $C_\rho(\mu,k,\eta)$ and $C_p(\mu,k,\eta)$ are explicit in appendix A. It is also useful to define here the two-point function
\be
\langle X^2 \rangle=\frac{1}{(2\pi)^3}\int\text{d}^3k \,|h_k|^2\, ,
\ee
which is also UV divergent and can be renormalized by means of the (extended) adiabatic regularization method, resulting in
\be
|h_k|^2_{\textrm{ren}}=|h_k|^2-C_{X^2}(\mu,k,\eta)\, .
\ee
The adiabatic counter-terms for the two point function are also given in appendix A. As a final comment, we remark that regularizing the stress-energy tensor requires adiabatic subtractions up to the 4th adiabatic order, while the two-point function only requires subtractions up to the 2nd order. As explained in \cite{Ferreiro:2022hik}, in order to construct the instantaneous vacuum state we choose a fixed $\mu=\mu_*$ which once fixed will not be modified in the subsequent calculations, therefore fixing all the possible renormalized couplings. 

\section{Instantaneous vacuum} \label{sec:IV}
In Ref. \cite{Agullo:2014ica} it was proposed a general method to construct a physically sensible vacuum state for free scalar fields in flat FLRW cosmologies. This method consists of choosing the vacuum, i.e., the initial conditions $\{h_k(\eta_0),h'_k(\eta_0)\}$, that make the renormalized vacuum expectation value of the stress-energy tensor vanish at a given time $\eta=\eta_0$ mode-by-mode, that is $\langle \rho_k\rangle(\eta_0)_{\textrm{ren}}=0$ and $\langle p_k\rangle(\eta_0)_{\textrm{ren}}=0$. In this first approach, the subtraction terms were generated using the standard adiabatic expansion.  By construction, this vacuum state is of adiabatic order four, ensuring the renormalizability of the stress-energy tensor at any time $\eta$. This proposal was later improved in Ref. \cite{Ferreiro:2022hik}, by including self-interactions and also the improved adiabatic renormalization scheme, briefly sketched in section \ref{sec:Yukawa}. In this section, we summarise the main points of this construction.\\

As previously mentioned, to choose a vacuum at $\eta=\eta_0$ implies to choose a particular set of initial conditions  $\{h_k(\eta_0),h'_k(\eta_0)\}$, that can be conveniently parametrized as
\bea \label{IV-1}
h_{ k}(\eta_0)=\frac{1}{a(\eta_0)\sqrt{2 W_{k}(\eta_0)}}\, , \qquad h'_{ k}(\eta_0)=\Big(-i W_{ k}(\eta_0)+\frac{V_{ k}(\eta_0)}{2}-\frac{a'(\eta_0)}{a(\eta_0)}\Big)h_{ k}(\eta_0)\, ,
\eea
where $V_k(\eta_0)$ has to be real and $W(\eta_0)$ has to be real and positive for all $k$ to ensure that the solutions are normalized. Furthermore, to guarantee that the stress-energy tensor is finite  after renormalization, $W_k(\eta_0)$ and $V_k(\eta_0)$ must obey the following asymptotic conditions
 \be \label{eq:initial-adiabaticCOND} W_k(\eta_0)=\Omega^{(4)}(\eta_0)+\mathcal{O}(k^{-4-\epsilon})\, , \quad V_k(\eta_0)=\frac{\partial_\eta \Omega^{(4)}}{\Omega^{(4)}}\Big|_{\eta_0}+\mathcal{O}(k^{-4-\epsilon})\, , \ee 
 where $\Omega^{(4)}$ refers to the adiabatic expansion up to 4th adiabatic order (for more details see Ref. \cite{Ferreiro:2022hik}). This condition is not enough to fix the vacuum univocally, and extra conditions have to be imposed. In this context, a physically reasonable condition is to impose that, at $\eta=\eta_0$, the renormalized stress-energy tensor should vanish mode-by-mode, namely
\bea \label{IV-cond}
\langle \rho_k \rangle (\eta_0)=C_\rho(\mu_{*}, k, \eta_0) \, ,\qquad \langle p _k \rangle(\eta_0) =C_p(\mu_{*}, k, \eta_0)\, .
\eea
Inserting the ansatz \eqref{IV-1} in \eqref{IV-cond} we arrive to
\bea \label{eq:IV-defW}
W_k(\eta_0)&=&\frac{2k^2+3\,Q (\eta_0) a(\eta_0)^2}{6a(\eta_0)^4(C_\rho(\mu_*,k,\eta_0)-C_p(\mu_*,k,\eta_0))} \label{W00}\, ,\\
\nonumber \\
V^{(\pm)}_k(\eta_0)&=&\frac{2a'(\eta_0)}{a(\eta_0)} \pm 2\sqrt{-W_k(\eta_0)^2-k^2  -a(\eta_0)^2Q(\eta_0)+4 a(\eta_0)^4 C_\rho(\mu_*,k,\eta_0)W_k(\eta_0)} \, . \label{V00}
\eea
It can be proved that $V_k^{(+)}$ is the appropriated solution for expanding universes, while $V_k^{(-)}$ should be used in the contracting case \cite{Agullo:2014ica}. As stressed above this solutions are consistent if and only if $W_k(\eta_0)\geq 0$ and $V_k(\eta_0)$ real. The second condition results in
\bea
\infty >r_k(\eta_0)=-W_k(\eta_0)^2-k^2  -a(\eta_0)^2 Q(\eta_0)+4 a(\eta_0)^4 C_\rho(\mu_*,k,\eta_0)W_k(\eta_0) \geq 0. \label{cond1}
\eea
In Ref. \cite{Ferreiro:2022hik} it was shown that the conditions above are not trivially satisfied for low values of $\mu_{*}$, and hence, a lower bound for this parameter $\mu_*\geq \mu_{\textrm{min}}$ has to imposed. It was also found that this lower bound is typically bigger than the effective mass $Q(\eta_0)$. For arbitrarily big values of $\mu_*$ the conditions are always satisfied, being $r_k\geq0$ the most restrictive one. On the other hand, it was additionally encountered that for big values of $\mu_*$ the initial backreaction effects, namely $\langle X^2 \rangle(\eta_0)_{\textrm{ren}}$, cannot be ignored, which is undesirable from the physical point of view. This fact effectively generates an upper bound for $\mu_*\leq\mu_{\textrm{max}}$.\\

Once the initial conditions are fixed using \eqref{W00} and \eqref{V00}, an unique solution to the mode equation \eqref{eq:mode-eq} is obtained. We call this solution the instantaneous vacuum. In the next section we introduce a second proposal for finding a preferred solution $h_k$ by looking  at the energy density. In analogy to instantaneous Hamiltonian diagonalization, the vacuum state can be defined as the state that minimizes the energy density but this time over a temporal window instead of at a given time $\eta=\eta_0$. This construction guarantees then that the resulting state is Hadamard, in contrast with the standard Hamiltonian diagonalization that can render ill defined vacua.

\section{States of Low Energy}
\label{sec:SLE}
In Ref. \cite{Olbermann:2007gn} it was proposed a very interesting method to construct Hadamard vacuum states in FLRW spacetimes. These states are defined by requiring that the vacuum expectation value of the energy density after averaging with a temporal window function $f^2$ is minimized. In this section, we briefly discuss this method for the problem under consideration: a scalar field with a Yukawa interaction. The vacuum expectation value of the smeared energy density for the mode $k$ is defined as
\be \label{eq:smearedED}
W_k[f] :=\int \text{d} \eta\, \sqrt{-g}\,  f^2(\eta) \, \langle \rho_k \rangle \, .
\ee
where $\langle \rho_k\rangle$ is given in \eqref{rhop}, $f^2$ is a positive definite window function, and $\sqrt{-g}=a^4$ is the 4-volume factor as introduced in \cite{Nadal23}. The main idea of this proposal is to find for which solution $h_k$ the smeared energy density  \eqref{eq:smearedED} is minimal. We note that since the subtraction terms are independent of the vacuum state, we can directly work with the formal, unrenormalized, result. In order to minimize $W_k$ it is very convenient to expand the scalar modes $h_k$ in terms of a basis of solutions as
\be \label{TS}h_k(\eta) = \lambda_k s_k(\eta) + \mu_k s_k^*(\eta) \ , \ee
where $\lambda_k$ and $\mu_k$ are complex numbers that must obey \be \label{eq:normSLE} |\lambda_k|^2 -|\mu_k|^2 =1\,.\ee
For our proposals, it is possible to choose $\mu_k$ real (see \cite{Olbermann:2007gn} for more details). The smeared energy density can be then written in terms of $\mu_k$ and $\lambda_k$ as
\be \label{eq:smearedc1c2}
W_k=(2 \mu_k^2+1)c_{k,1}+2 \mu_k \textrm{Re}(\lambda_k c_{k,2})\, ,
\ee
where we have defined ($c_1\equiv c_{k,1}$, $c_2\equiv c_{k,2}$)
\bea \label{eq:c1}
&&c_1=\frac{1}{2} \int  \text{d} \eta \,a^{2}\, f^2\,\Big(|s'_k|^2+(k^2+a^2 g a^2 \phi^2)|s_k|^2\Big) \, ,\\
&&c_2=\frac{1}{2} \int  \text{d} \eta \, a^2\,f^2\left(s'_k{}^2+(k^2 +g a^2 \phi^2)^2 s_k^2\right)\, . \label{eq:c2}
\eea
We do only consider here the cases with $c_1$ positive for all $k$. In this context, and assuming that $\mu_k$ is negative, it can be shown that \eqref{eq:smearedc1c2} is minimized for 
\be\label{mu-lambda}  \lambda_k=e^{-i(\text{Arg} \ c_2)}  \sqrt{\frac{c_1}{2\sqrt{c_1^2 -|c_2|^2}} +\frac{1}{2}} \ ; \qquad \mu_k= -\sqrt{\frac{c_1}{2\sqrt{c_1^2 -|c_2|^2}} -\frac{1}{2}} \, . \ee
In other words, the state \eqref{TS} with $\mu_k$ and $\lambda_k$ given in \eqref{mu-lambda} is the state that minimizes the smeared energy density over the window function $f^2$. The resulting state is independent of the fiducial solution $s_k$. \\

A natural question appears: how different is this construction from the instantaneous vacuum defined in the previous section?  From the formal side, perhaps the most important difference is that the states of low energy described here are states of infinite adiabatic order (i.e., Hadamard states), while the instantaneous vacuum is a state of fourth adiabatic order. In practice, however, this difference could be very small, and the physical predictions for each vacuum could be almost the same if we focus only on the construction of the two point function and the stress-energy tensor. On the other hand, it is also important to stress that the instantaneous vacuum is defined in terms of the subtraction terms so it can be constructed even for the cases where analytical solutions to the mode equations $s_k$ are not available. This makes the instantaneous vacuum very convenient for numerical purposes. \\
 
 In the next section we will compare the instantaneous vacuum defined at $\eta=\eta_0$ with the state of low energy for a window function centered at $\eta=\eta_0$. For simplicity we will choose a Gaussian window function
  \be\label{eq:Gaussian}
f^2= \frac{1}{\sqrt{\pi}\epsilon}\, e^{{-\frac{(\eta-\eta_0)^2}{\epsilon^2}}}\, .\ee


\section{States of Low Energy vs. Instantaneous vacuum}
\label{sec:minkowski}

For simplicity, let us restrict to Minkowski spacetime. First, we will compute the vacuum state using the method described above and then we will compare it with the instantaneous vacuum  at $\eta=\eta_0$. 
Ignoring backreaction effects, the equation for the background takes a very simple form 
\be
\phi''+M^2\phi=0\, .
\ee
 The solution to this equation for the initial conditions $\phi(\eta_0)=\phi_0$ and $\phi'(\eta_0)=0$ is \be \phi=\phi_0\cos[M(\eta-\eta_0)]\, .\ee 
The equation for the field modes reads
\be\label{eq:modes-d-mink}
h_k''+\left(k^2+g\phi_0^2\cos^2[M(\eta-\eta_0)]\right)\,h_k=0\, , 
\ee
The solution to the equation above can be written in terms of the Mathieu functions $C(a,b;x)$ and $S(a,b;x)$ (see Ref. \cite{WolframMathieu}).  For our proposals, and following section \ref{sec:SLE}, it is convenient to express the solution for $h_k$ as 
\be
h_k=\lambda_k \, s_k(\eta)+ \mu_k \, s_k^*(\eta)\,.
\ee
where
\be
s_k(\eta)=\frac{1}{\sqrt{M}\sqrt{2\kappa}} \left(\frac{C(2q+\kappa^2,-q;M(\eta-\eta_0))}{C(2q+\kappa^2,-q;0)}-iM \frac{S(2q+\kappa^2,-q;M(\eta-\eta_0))}{S'(2q+\kappa^2,-q;0)}\right)\, .
\ee
with $q:= \frac{g\phi_0^2}{4M^4}$ and $\kappa=\frac{k}{M}$. It is important to stress that the fiducial solutions $\{s_k(\eta),s^{*}_k(\eta)\}$ satisfy the Wronskian condition, i.e., $s_k s_k^{* \prime}-s_k^* s_k^{\prime}=i$, so that 
  the constants $\lambda_k$ and $\mu_k$ satisfy the normalization condition given in Eq.  \eqref{eq:normSLE}. We now proceed to compute the state of low energy centered at $\eta=\eta_0$ using the Gaussian window function defined in \eqref{eq:Gaussian}. \\

To this end, we first compute 
$c_1$ and $c_2$ using Eqs. \eqref{eq:c1} and  \eqref{eq:c2}. The $\eta$-integrals can be done numerically with the help of  the {\it Mathematica} software. It is not difficult to see that
\be
\textrm{Im}\, c_{2}=0\, \quad \text{and} \quad c_{2}>0\, . 
\ee
Therefore the phase of $\lambda_k$ is $\textrm{Arg} \lambda_k=-\textrm{Arg} c_2=0$, and we can conclude that $\mu_k$ and $\lambda_k$ are both real numbers. Using Eq. \eqref{mu-lambda} we then compute numerically $\lambda_k$ and $\mu_k$.\\ 




In summary, the solution
\be
h_k(\eta)=\lambda_k\, s_k(\eta) + \mu_k \,s_k^*(\eta)\, ,
\ee
with the constants $\lambda_k$ and $\mu_k$ obtained above is the state of low energy centered at $\eta=\eta_0$ obtained with a Gaussian smearing function. The now goal is to compare this state with the instantaneous vacuum. To this end, we can evaluate both solutions at $\eta=\eta_0$. On the one hand, the state of low energy at $\eta=\eta_0$ reads  
\be \label{eq:chiz0SLE}h_k(\eta_0)=\frac{\lambda_k+\mu_k}{\sqrt{M}\sqrt{2\kappa}}\,\, , \qquad h_k'(\eta_0)=\frac{-i\sqrt{M}\sqrt{\kappa}}{\sqrt{2}}(\lambda_k-\mu_k)\, . \ee
On the other hand, for the instantaneous vacuum we have
\bea
{W}_k\left(\eta_0\right)&=&\frac{2  \kappa^2+3 Q_*\left(\eta_0\right)}{6 \left(C_\rho\left({\mu}_*, \kappa, \eta_0\right)-C_p\left({\mu}_*, \kappa, \eta_0\right)\right)}\, ,\\
M^{-1} V^{(\pm) }_k(\eta_0)&=& \pm 2\sqrt{-M^{-2} W_k(\eta_0)^2-\kappa^2 -Q_*(\eta_0)+4  M^{-2} C_\rho(\mu_*,\kappa,\eta_0) W_k(\eta_0)}\, .
\eea
 where $Q_*=4 q \cos^2(M(\eta-\eta_0))$. A detailed analysis of the instantaneous vacuum for this case including the choice of the $\mu_*$ scale can be found in Ref. \cite{Ferreiro:2022hik}. It can be argued that the choice 
\be 
\frac{{\mu}_*^2}{M^{2}}={4}q+\sqrt{\tfrac{2}{3}} \sqrt{4q} \,.\label{mup}
\ee
could be somewhat favoured since it minimizes the initial backreaction effects. More specifically, for this value of $\mu_*$ we get $\langle X^2\rangle_{\textrm{ren}}(\eta_0)=0$.\\

\begin{figure}[h]
    \centering \hspace{-0.5cm}
    \includegraphics[width=7.7cm]{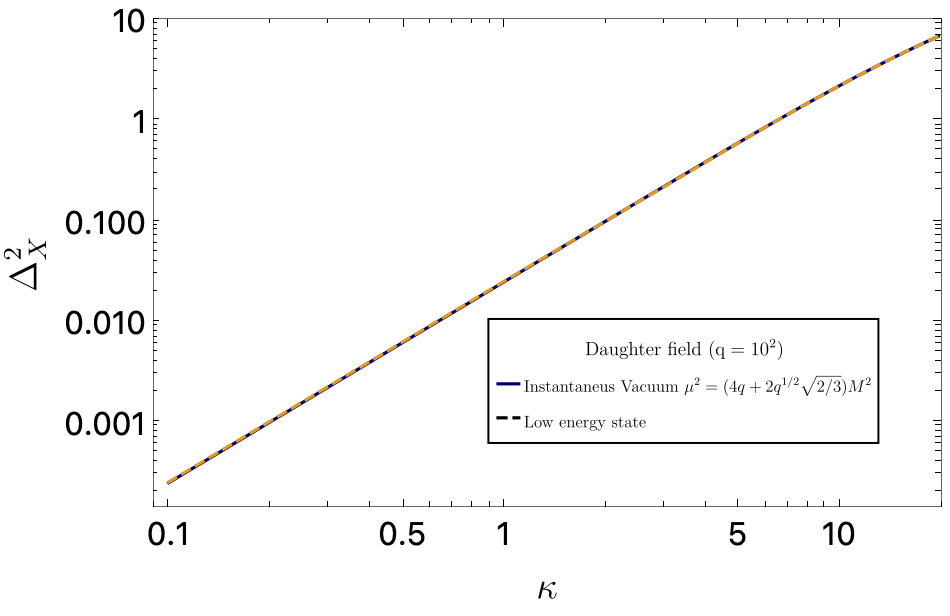}  \includegraphics[width=7.7cm]{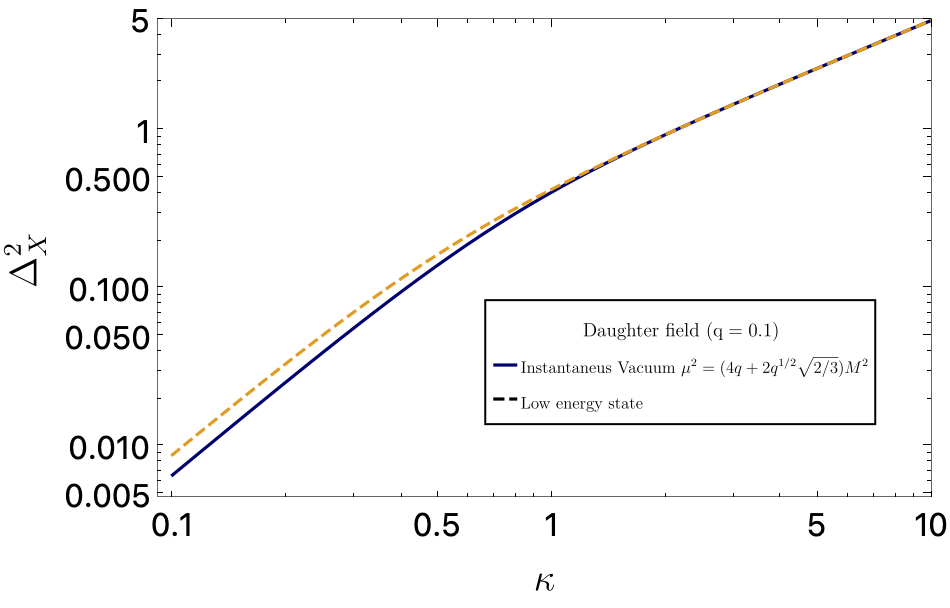}
    \caption{{\small We plot the spectrum of the two-point function for two different values of $q$.} }
    \label{fig:1}
\end{figure}

\begin{figure}[h]
    \centering
    \includegraphics[width=7.5cm]{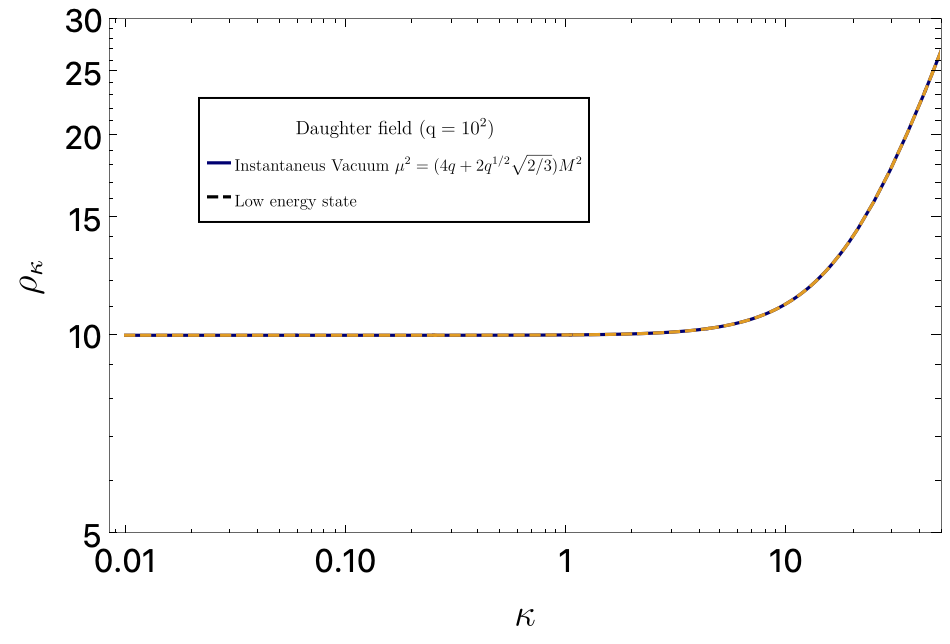}  \includegraphics[width=7.5cm]{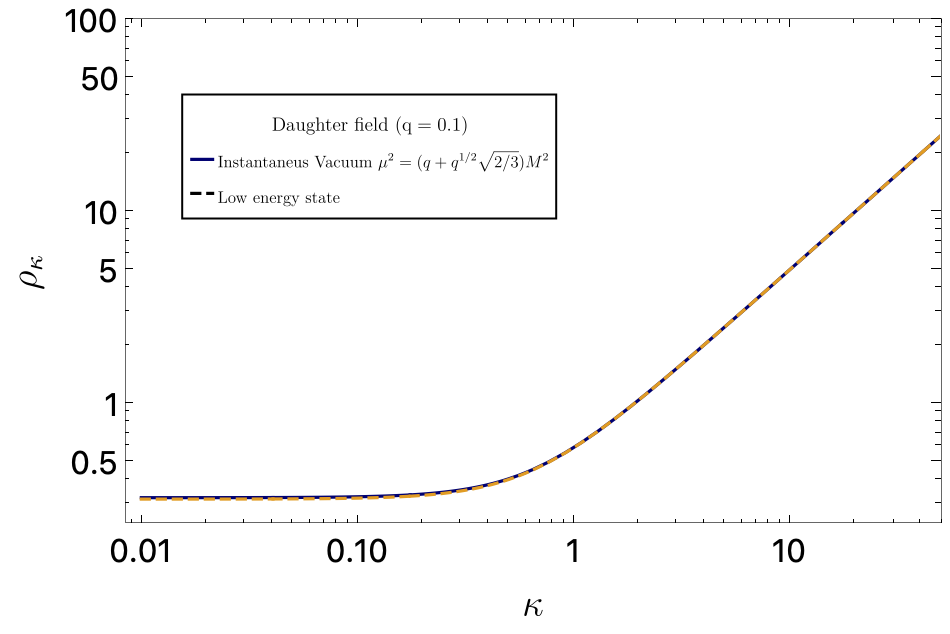}
    \caption{{\small We plot the spectrum of the energy density for two different values of $q$..} }
    \label{fig:2}
\end{figure}

Using this results we can finally compare the energy density $\langle \rho_k\rangle(\eta_0)$ and the two-point function $|h_k|^2(\eta_0)$ for these two vacua. We fix $M \epsilon=1$ in the smearing function. For  $\mu_*$, we choose the preferred value given in Eq. \eqref{mup}. In figures \ref{fig:1} and \ref{fig:2} we plot both the the spectrum of the two point function, i.e. $\Delta_X^2:=k^2|h_k|^2$, and the energy density defined in \eqref{rhop} for the instantaneous vacuum and the low energy states at $\eta=\eta_0$ for two different values of $q$. 


We observe that for the cases under consideration the instantaneous vacuum and the state of low energy are almost equivalent. This equivalence appears to be bigger for higher values of the Yukawa coupling $q$.

\section{Conclusions} \label{sec:conclusions}

The choice of the vacuum state in cosmology is not unique, and additional criteria must be applied to establish a preferred vacuum for the theory. Our work examined two distinct possibilities achieved by imposing different conditions on the stress-energy tensor. \\

On the one hand, the instantaneous vacuum is defined by requiring the renormalized stress-energy tensor to vanish at a given initial time $t=t_0$ and at a given energy scale $\mu=\mu_*$. This state is entirely determined by the mode-by-mode subtraction terms, and guarantees a finite vacuum expectation value of the stress energy tensor at any time, which makes it very convenient for numerical computations. This is a fourth adiabatic order vacuum and therefore it does not satify the Hadamard condition.\\

On the other hand, the states of low energy are determined by requiring the vacuum expectation value of the energy density to be minimal when averaged over a time-window function $f^2$ with support around a given time $t=t_0$. By construction, these states satisfy the Hadamard condition. In practice, however, these states are not easy to implement, since analytical solutions to the scalar field mode equations are required to obtain the desired vacua.  \\

We have explicitly compared these two possibilities for a quantized scalar field with a Yukawa interaction in Minkowski spacetimes. We observe that for the cases under consideration these two vacuum states are almost equivalent. We have explicitly compared the spectrum of the two-point function and the energy density at the initial time $\eta=\eta_0$, obtaining the same infrared behaviour for both choices. 

\section*{Acknowledgements}

We thank J. Navarro-Salas and F. Torrentí for very useful discussions.  We also acknowledge the organisers of the conference ``Avenues in Quantum field Theory in Curved
Spacetimes 2022” for a very interesting  meeting, where an earlier version of this work was presented. A.F.~is supported by the Margarita Salas fellowship MS21-085 of the University of Valencia. S.P.~is supported by the Leverhulme Trust, Grant No.~RPG-2021-299.

\appendix
\section{Adiabatic subtractions for $C_{X^2}$, $C_\rho$ and $C_p$}
Defining $s=Q-\mu^2$ and using $\omega=\sqrt{k^2+a^2 \mu^2}$, the adiabatic subtractions for the two-point function $|h_k|^2$, the energy density, and the pressure read
\be
C_{X^2}=\frac{1}{2a^2 \omega}-\frac{3(\omega')^2}{16 a^3 \omega^5}+\frac{a''}{4 a^3 \omega^3}+\frac{\omega''}{8 a^2 \omega^4}-\frac{s}{4 \omega^3}\, .
\ee
\bea
C_\rho&=&\frac{\omega }{2 a^4}+\frac{s}{4 a^2 \omega }+\frac{\left(a'\right)^2}{4
   a^6 \omega }-\frac{3 \xi  \left(a'\right)^2}{2 a^6 \omega }+\frac{a'
   \omega '}{4 a^5 \omega ^2}-\frac{3 \xi  a' \omega '}{2 a^5 \omega
   ^2}+\frac{\left(\omega '\right)^2}{16 a^4 \omega ^3}-\frac{s \sigma }{8 a^2 \omega ^3}+\frac{\sigma ^2}{16 a^4 \omega
   ^3}\\
   &&-\frac{\sigma  \left(a'\right)^2}{8 a^6 \omega ^3}+\frac{3 \xi 
   \sigma  \left(a'\right)^2}{4 a^6 \omega ^3}+\frac{a' \sigma '}{8 a^5
   \omega ^3}-\frac{3 \xi  a' \sigma '}{4 a^5 \omega ^3}
   -\frac{3 \sigma 
   a' \omega '}{8 a^5 \omega ^4}+\frac{9 \xi  \sigma  a' \omega '}{4 a^5
   \omega ^4}+\frac{\sigma ' \omega '}{16 a^4 \omega ^4}-\frac{3 s
   \left(\omega '\right)^2}{32 a^2 \omega ^5}\nonumber\\
   &&-\frac{\sigma  \left(\omega
   '\right)^2}{16 a^4 \omega ^5}-\frac{3 \left(a'\right)^2 \left(\omega
   '\right)^2}{32 a^6 \omega ^5}+\frac{9 \xi  \left(a'\right)^2
   \left(\omega '\right)^2}{16 a^6 \omega ^5}-\frac{15 a' \left(\omega
   '\right)^3}{32 a^5 \omega ^6}+\frac{45 \xi  a' \left(\omega
   '\right)^3}{16 a^5 \omega ^6}-\frac{45 \left(\omega '\right)^4}{256 a^4
   \omega ^7}\nonumber\\
   &&+\frac{s \omega ''}{16 a^2 \omega ^4}-\frac{\sigma  \omega
   ''}{16 a^4 \omega ^4}+\frac{\left(a'\right)^2 \omega ''}{16 a^6 \omega
   ^4}-\frac{3 \xi  \left(a'\right)^2 \omega ''}{8 a^6 \omega ^4}+\frac{7
   a' \omega ' \omega ''}{16 a^5 \omega ^5}-\frac{21 \xi  a' \omega '
   \omega ''}{8 a^5 \omega ^5}+\frac{5 \left(\omega '\right)^2 \omega
   ''}{32 a^4 \omega ^6}\nonumber\\
   &&+\frac{\left(\omega ''\right)^2}{64 a^4 \omega
   ^5}-\frac{a' \omega ^{(3)}}{16 a^5 \omega ^4}+\frac{3 \xi  a' \omega
   ^{(3)}}{8 a^5 \omega ^4}-\frac{\omega ' \omega ^{(3)}}{32 a^4 \omega
   ^5}\, .\nonumber
\eea
\bea
C_p&=&\frac{k^2}{6 a^4 \omega }+\frac{\mu ^2 \sigma }{12 a^2 \omega
   ^3}-\frac{s}{4 a^2 \omega }+\frac{s \xi }{a^2 \omega }+\frac{\sigma }{6
   a^4 \omega }-\frac{\xi  \sigma }{a^4 \omega
   }+\frac{\left(a'\right)^2}{4 a^6 \omega }-\frac{3 \xi 
   \left(a'\right)^2}{2 a^6 \omega }+\frac{a' \omega '}{4 a^5 \omega
   ^2}\\
   &&-\frac{3 \xi  a' \omega '}{2 a^5 \omega ^2}+\frac{\mu ^2
   \left(\omega '\right)^2}{16 a^2 \omega ^5}+\frac{3 \left(\omega
   '\right)^2}{16 a^4 \omega ^3}-\frac{\xi  \left(\omega '\right)^2}{a^4
   \omega ^3}-\frac{\xi  a''}{a^5 \omega }+\frac{6 \xi ^2 a''}{a^5 \omega
   }-\frac{\mu ^2 \omega ''}{24 a^2 \omega ^4}-\frac{\omega ''}{12 a^4
   \omega ^2}\nonumber\\
   &&+\frac{\xi  \omega ''}{2 a^4 \omega ^2}-\frac{\mu ^2 \sigma ^2}{16 a^2 \omega ^5}+\frac{s \sigma }{8 a^2 \omega^3}-\frac{s \xi  \sigma }{2 a^2 \omega ^3}-\frac{\sigma ^2}{16 a^4
   \omega ^3}+\frac{\xi  \sigma ^2}{2 a^4 \omega ^3}-\frac{\sigma 
   \left(a'\right)^2}{8 a^6 \omega ^3}+\frac{3 \xi  \sigma 
   \left(a'\right)^2}{4 a^6 \omega ^3}\nonumber\\
   &&+\frac{a' \sigma '}{8 a^5 \omega
   ^3}-\frac{3 \xi  a' \sigma '}{4 a^5 \omega ^3}-\frac{3 \sigma  a'
   \omega '}{8 a^5 \omega ^4}+\frac{9 \xi  \sigma  a' \omega '}{4 a^5
   \omega ^4}+\frac{5 \mu ^2 \sigma ' \omega '}{48 a^2 \omega ^6}+\frac{13
   \sigma ' \omega '}{48 a^4 \omega ^4}-\frac{3 \xi  \sigma ' \omega '}{2
   a^4 \omega ^4}+\frac{3 s \left(\omega '\right)^2}{32 a^2 \omega ^5}\nonumber\\
   &&-\frac{25 \mu ^2 \sigma  \left(\omega '\right)^2}{96 a^2
   \omega ^7}-\frac{3
   s \xi  \left(\omega '\right)^2}{8 a^2 \omega ^5}-\frac{7 \sigma 
   \left(\omega '\right)^2}{12 a^4 \omega ^5}+\frac{27 \xi  \sigma 
   \left(\omega '\right)^2}{8 a^4 \omega ^5}-\frac{3 \left(a'\right)^2
   \left(\omega '\right)^2}{32 a^6 \omega ^5}+\frac{9 \xi 
   \left(a'\right)^2 \left(\omega '\right)^2}{16 a^6 \omega ^5}\nonumber\\
   &&-\frac{15
   a' \left(\omega '\right)^3}{32 a^5 \omega ^6}+\frac{45 \xi  a'
   \left(\omega '\right)^3}{16 a^5 \omega ^6}-\frac{105 \mu ^2
   \left(\omega '\right)^4}{256 a^2 \omega ^9}-\frac{255 \left(\omega
   '\right)^4}{256 a^4 \omega ^7}+\frac{45 \xi  \left(\omega '\right)^4}{8
   a^4 \omega ^7}+\frac{\xi  \sigma  a''}{2 a^5 \omega ^3}\nonumber\\
   &&-\frac{3 \xi ^2
   \sigma  a''}{a^5 \omega ^3}+\frac{3 \xi  \left(\omega '\right)^2 a''}{8
   a^5 \omega ^5}-\frac{9 \xi ^2 \left(\omega '\right)^2 a''}{4 a^5 \omega
   ^5}-\frac{\mu ^2 \sigma ''}{48 a^2 \omega ^5}-\frac{\sigma ''}{24 a^4
   \omega ^3}+\frac{\xi  \sigma ''}{4 a^4 \omega ^3}+\frac{5 \mu ^2 \sigma
    \omega ''}{48 a^2 \omega ^6}\nonumber\\
   &&-\frac{s \omega ''}{16 a^2 \omega
   ^4}+\frac{s \xi  \omega ''}{4 a^2 \omega ^4}+\frac{7 \sigma  \omega
   ''}{48 a^4 \omega ^4}-\frac{\xi  \sigma  \omega ''}{a^4 \omega
   ^4}+\frac{\left(a'\right)^2 \omega ''}{16 a^6 \omega ^4}-\frac{3 \xi 
   \left(a'\right)^2 \omega ''}{8 a^6 \omega ^4}+\frac{7 a' \omega '
   \omega ''}{16 a^5 \omega ^5}\nonumber\\
   &&-\frac{21 \xi  a' \omega ' \omega ''}{8 a^5
   \omega ^5}+\frac{35 \mu ^2 \left(\omega '\right)^2 \omega ''}{64 a^2
   \omega ^8}+\frac{5 \left(\omega '\right)^2 \omega ''}{4 a^4 \omega
   ^6}-\frac{115 \xi  \left(\omega '\right)^2 \omega ''}{16 a^4 \omega
   ^6}-\frac{\xi  a'' \omega ''}{4 a^5 \omega ^4}+\frac{3 \xi ^2 a''
   \omega ''}{2 a^5 \omega ^4}\nonumber\\
   &&-\frac{5 \mu ^2 \left(\omega ''\right)^2}{64
   a^2 \omega ^7}-\frac{9 \left(\omega ''\right)^2}{64 a^4 \omega
   ^5}+\frac{7 \xi  \left(\omega ''\right)^2}{8 a^4 \omega ^5}-\frac{a'
   \omega ^{(3)}}{16 a^5 \omega ^4}+\frac{3 \xi  a' \omega ^{(3)}}{8 a^5
   \omega ^4}-\frac{5 \mu ^2 \omega ' \omega ^{(3)}}{48 a^2 \omega
   ^7}\nonumber\\
   &&-\frac{23 \omega ' \omega ^{(3)}}{96 a^4 \omega ^5}+\frac{11 \xi 
   \omega ' \omega ^{(3)}}{8 a^4 \omega ^5}+\frac{\mu ^2 \omega ^{(4)}}{96
   a^2 \omega ^6}+\frac{\omega ^{(4)}}{48 a^4 \omega ^4}-\frac{\xi  \omega
   ^{(4)}}{8 a^4 \omega ^4}\, . \nonumber
\eea



\section*{References}
\bibliography{iopart-num}

\providecommand{\newblock}{}
\begin{thebibliography}{10}
\expandafter\ifx\csname url\endcsname\relax
  \def\url#1{{\tt #1}}\fi
\expandafter\ifx\csname urlprefix\endcsname\relax\def\urlprefix{URL }\fi
\providecommand{\eprint}[2][]{\url{#2}}

\bibitem{Ferreiro:2022hik}
Ferreiro A and Pla S 2022 {\em Phys. Rev. D\/} {\bf 106} 065015
  (\textit{Preprint} \eprint{2206.08200})

\bibitem{Wald94}
Wald R~M 1995 {\em {Quantum Field Theory in Curved Space-Time and Black Hole
  Thermodynamics}\/} Chicago Lectures in Physics (Chicago, IL: University of
  Chicago Press) ISBN 978-0-226-87027-4

\bibitem{Fewster13}
Fewster C~J and Verch R 2013 {\em Class. Quant. Grav.\/} {\bf 30} 235027
  (\textit{Preprint} \eprint{1307.5242})

\bibitem{BF2000}
Brunetti R and Fredenhagen K 2000 {\em Commun. Math. Phys.\/} {\bf 208}
  623--661 (\textit{Preprint} \eprint{math-ph/9903028})

\bibitem{hollands-wald-01}
Hollands S and Wald R~M 2001 {\em Commun. Math. Phys.\/} {\bf 223} 289--326
  (\textit{Preprint} \eprint{gr-qc/0103074})

\bibitem{hollands-wald-02}
Hollands S and Wald R~M 2002 {\em Commun. Math. Phys.\/} {\bf 231} 309--345
  (\textit{Preprint} \eprint{gr-qc/0111108})

\bibitem{parker-toms}
Parker L~E and Toms D 2009 {\em {Quantum Field Theory in Curved Spacetime}:
  {Quantized Field and Gravity}\/} Cambridge Monographs on Mathematical Physics
  (Cambridge University Press) ISBN 978-0-521-87787-9

\bibitem{Fulling}
Fulling S~A 1989 {\em {Aspects of Quantum Field Theory in Curved Space-time}\/}
  vol~17 (Cambridge University Press)

\bibitem{Pirk93}
Pirk K~T 1993 {\em Phys. Rev. D\/} {\bf 48} 3779--3783 (\textit{Preprint}
  \eprint{gr-qc/9211003})

\bibitem{Hollands01}
Hollands S 2001 {\em Commun. Math. Phys.\/} {\bf 216} 635--661
  (\textit{Preprint} \eprint{gr-qc/9906076})

\bibitem{Olbermann:2007gn}
Olbermann H 2007 {\em Class. Quant. Grav.\/} {\bf 24} 5011--5030
  (\textit{Preprint} \eprint{0704.2986})

\bibitem{Martin-Benito:2021szh}
Mart\'\i{}n-Benito M, Neves R~B and Olmedo J 2021 {\em Phys. Rev. D\/} {\bf
  103} 123524 (\textit{Preprint} \eprint{2104.03035})

\bibitem{Nadal23}
Nadal-Gisbert S, Navarro-Salas J and Pla S 2023 {Low Energy States and CPT
  invariance at the Big Bang} (\textit{Preprint} \eprint{2302.08812})

\bibitem{Alvarez-Dominguez:2023ten}
\'Alvarez-Dom\'\i{}nguez A, Garay L~J, Mart\'\i{}n-Benito M and Neves R~B 2023
  {States of low energy in the Schwinger effect} (\textit{Preprint}
  \eprint{2303.15294})

\bibitem{Agullo:2014ica}
Agullo I, Nelson W and Ashtekar A 2015 {\em Phys. Rev. D\/} {\bf 91} 064051
  (\textit{Preprint} \eprint{1412.3524})

\bibitem{Ferreiro:2018oxx}
Ferreiro A and Navarro-Salas J 2019 {\em Phys. Lett. B\/} {\bf 792} 81--85
  (\textit{Preprint} \eprint{1812.05564})

\bibitem{WolframMathieu}
Wolfram-Research I {\em Wolfram Language and System Documentation Center,
  Mathieu and Related Functions\/} (Champaign, IL, 2022)

\end{thebibliography}

\end{document}